\documentclass[12pt]{article}

\usepackage{times}
\usepackage{geometry}
\usepackage[utf8]{inputenc}
\usepackage{enumitem,amssymb}
\usepackage{ragged2e}
\usepackage{fancyhdr}
\usepackage[sort,numbers]{natbib}
\usepackage{graphics,graphicx}
\usepackage{color}
\usepackage{xcolor}
\usepackage{lineno}
\usepackage{geometry}
\usepackage{wasysym}
\usepackage{csquotes}

\usepackage[compact]{titlesec}

\usepackage{rotating}
\usepackage{pdflscape}

\newlist{thematic}{itemize}{8}
\setlist[thematic]{label=$\square$}
\usepackage{pifont}
\newcommand{\cmark}{\ding{51}}%
\newcommand{\done}{\rlap{$\square$}{\raisebox{2pt}{\large\hspace{1pt}\cmark}}%
\hspace{-2.5pt}}

\begin{document}
\raggedright
\huge
Astro2020 Science White Paper \linebreak

Gamma Rays and Gravitational Waves \linebreak
\normalsize

\noindent \textbf{Thematic Areas:} \hspace*{60pt} $\square$ Planetary Systems \hspace*{10pt} $\square$ Star and Planet Formation \hspace*{20pt}\linebreak
$\done$ Formation and Evolution of Compact Objects \hspace*{31pt} $\done$ Cosmology and Fundamental Physics \linebreak
  $\square$  Stars and Stellar Evolution \hspace*{1pt} $\square$ Resolved Stellar Populations and their Environments \hspace*{40pt} \linebreak
  $\square$    Galaxy Evolution   \hspace*{45pt} $\done$             Multi-Messenger Astronomy and Astrophysics \hspace*{65pt} \linebreak
  
\textbf{Principal Author:}

Name:	Eric Burns
 \linebreak						
Institution:  NASA Goddard
 \linebreak
Email: eric.burns@nasa.gov
 \linebreak
Phone:  +1-301-286-4664
 \linebreak

\noindent \textbf{Co-authors:} 

\noindent S. Zhu (Albert Einstein Institute), C. M. Hui (NASA Marshall), S. Ansoldi (Universita` di Udine), S. Barthelmy (NASA Goddard), S. Boggs (UCSD), S. B. Cenko (NASA Goddard), N. Christensen (Observatoire de la C{\^o}te d'Azur), C. Fryer (LANL), A. Goldstein (USRA), A. Harding (NASA Goddard), D. Hartmann (Clemson University), A. Joens (George Washington University), G. Kanbach (Max Planck Institute for extraterrestrial Physics), M. Kerr (NRL), C. Kierans (NASA Goddard), J. McEnery (NASA Goddard), B. Patricelli (INFN), J. Perkins (NASA Goddard), J. Racusin (NASA Goddard), P. Ray (NRL), J. Schlieder (NASA Goddard), H. Schoorlemmer (MPI-HD), F. Sch{\"u}ssler (CEA Irfu DPhP), A. Stamerra (INAF), J. Tomsick (UC Berkeley, SSL), Z. Wadiasingh (NASA Goddard), C. Wilson-Hodge (NASA Marshall), G. Younes (George Washington University), B. Zhang (University of Nevada Las Vegas), A. Zoglauer (UC Berkeley, SSL)
  \justify
  

\noindent \textbf{Abstract:}

\noindent The first multimessenger observation of a neutron star merger was independently detected in $\gamma$-rays by Fermi-GBM and INTEGRAL SPI-ACS and gravitational waves by Advanced LIGO and Advanced Virgo. Gravitational waves are emitted from systems with accelerating quadrupole moments, and detectable sources are expected to be compact objects. Nearly all distant astrophysical $\gamma$-ray sources are compact objects. Therefore, serendipitous observations of these two messengers will continue to uncover the sources of gravitational waves and $\gamma$-rays, and enable multimessenger science across the Astro2020 thematic areas. This requires upgrades to the ground-based gravitational wave network and $\sim$keV-MeV $\gamma$-ray coverage for observations of neutron star mergers, and broadband coverage in both gravitational waves and $\gamma$-rays to monitor other expected joint sources.

\pagebreak
\section{Introduction}
The joint detection of $\gamma$-rays and gravitational waves (GWs) from merging neutron stars (NSs) ushered in a new era of multimessenger astronomy \citep{GW170817_LVC, GW170817_GBM, GW170817-GRB170817A, GW170817_MMAD}. This event has led the community to produce on average $>$3 papers/day. NS mergers will continue to be the canonical multimessenger source for the foreseeable future as the joint detections of these events become more common, but they are not the only expected sources of GWs. We summarize here the state of future GW and $\gamma$-ray observatories, the potential multimessenger sources, the science they enable, and the science that is achievable in a given mission size.

\subsection{Gravitational Waves}
GWs were first conceptualized by Poincar{\'e} \citep{GW_Poincare} and first predicted in the context of General Relativity (GR) in 1916 \citep{Einstein_1916,einstein_GWs_1918}. Their existence was indirectly confirmed by measuring the orbital frequency evolution of the Hulse-Taylor pulsar \citep{HTP_GWs_1982_nobel}, they were directly observed only a few years ago \citep{GW150914}, and we now have the first GW catalog \citep{LVC_GWTC1}. The strongest GWs detected so far have caused changes in the relative length of spacetime at Earth on the order of $10^{-21}$. The most sensitive GW detectors are interferometers and generally provide all-sky coverage.  A summary of facilities and their capabilities is given in Table \ref{tab:GW_network}.

\begin{table}[!b]
\centering
\begin{tabular}{|c|c|c|c|c|c|}
\hline
Instrument         & Year        & Frequency Range   & BNS Range     & BNS Rates (1/year)   & Ref. \\
\hline
\hline
GEO600             & 1995-       & $\sim$150-3000 Hz  &  &  & \citep{GEO600}\\
Advanced LIGO      & 2015-       & $\sim$20-1000 Hz   & 173 Mpc       & 0 (O1; 2015-2016)        &      \citep{Advanced_LIGO,LIGO_173Mpc}    \\
Advanced Virgo     & 2016-       & $\sim$20-1000 Hz   & 125 Mpc       & 1 (O2; 2017-2018)        &      \citep{Advanced_Virgo}        \\
KAGRA              & 2019+       & $\sim$20-700 Hz   & 140 Mpc       & 4-80 (2020+)     &  \citep{KAGRA}    \\
LIGO-India         & 2024+       & $\sim$20-1000 Hz   & 173 Mpc       & 11-180 (2024+)   &   \citep{review_future_GW_network_2018}   \\
\hline
Advanced LIGO+     & 2025+       & $\sim$20-1000 Hz  & 325 Mpc       & \textgreater{}100  &  \citep{LIGO_Aplus}  \\
Advanced Virgo+    & 2025+       & $\sim$20-1000 Hz  & 215 Mpc       &                   &    \\
\hline
LIGO Voyager       & 2028+ & $\sim$10-5,000 Hz & $\sim$1 Gpc   & \textgreater{}1,000  &  \citep{LIGO_Voyager} \\
\hline
Einstein Telescope & 2035+ & 1-10,000 Hz       & $\sim$few Gpc & \textgreater{}10,000 & \citep{GW_IFO_3rd_Gen_EU} \\
Cosmic Explorer    & 2035+ & 5-10,000 Hz       & $\sim$few Gpc &    & \citep{GW_IFO_3rd_Gen_US}  \\               
\hline
\hline
LISA & 2034+ & 0.1-100 mHz & & & \citep{LISA} \\
\hline
\hline
PTAs & 2005- & 1-1000 nHz & & & \citep{PTAs} \\
\hline
\end{tabular}
\caption{A summary of the active (GEO600 through Advanced Virgo), funded (GEO600 through Advanced LIGO+), and proposed interferometers. The first block consists of the $\sim$Hz-kHz ground-based interferometer network. Their sensitivity is usually quoted by the binary neutron star (BNS) range (the sky and orientation average detection distance); we expect vast detection rate increases in the coming years. The second block consists of LISA, a space-based interferometer sensitive in the $\sim$mHz regime. The last block is for Pulsar Timing Arrays (PTAs) that use observations of well-behaved pulsars as effective interferometer arms, sensitive to the $\sim$nHz regime.}\label{tab:GW_network}
\end{table}

Searches for GW signals in LIGO and Virgo can be classified according to their methodology, based on the type of GW emission. These classifications roughly extend to other GW detectors as well (note that we will not discuss searches for the stochastic GW background).
\textbf{Compact Binary Coalescences (CBCs)} are the mergers of compact objects (generally, black holes (BHs) and NSs), have GR-predicted waveforms, and encompass all GW detections so far. Searches for \textbf{Continuous Waves (CWs)} look for persistent sources of GW emission with waveforms predicted by GR. The signals are approximately monochromatic, evolve very slowly, and come from rotating non-axisymmetric systems. GW \textbf{Burst} searches are sensitive to transients with unmodeled or unspecified waveforms. \textbf{Intermediate duration} GWs have timescales in between those of CBC/Burst and CW signals, and are a comparatively new class. Searches for intermediate duration GWs either extend Burst searches to longer timescales or modify CW searches to shorter timescales and more rapid evolution.

\subsection{Gamma rays}
$\gamma$-rays are the most energetic form of light. In astrophysics the term is not restricted to photons from nuclear processes; as such, the lower limit of what constitutes $\gamma$-rays is somewhat fuzzy, generally set between $\sim$10-100 keV with the overlap sometimes referred to as hard X-rays or soft $\gamma$-rays. The upper limit is unbounded. $\gamma$-rays cover as many orders of magnitude in energy as all other observed light combined. 

A summary of representative $\gamma$-ray observatories is given in Table \ref{tab:gamma-rays}.
$\gamma$-rays between $\sim$keV-GeV energies can only be observed from space. These instruments are broadly classed as scintillators, coded masks \citep[e.g.][]{caroli1987coded}, Compton \citep[e.g.][]{von1989imaging}, and pair-conversion telescopes. Higher energy photons are observed indirectly through Cherenkov radiation as the photons pass through water (in enclosed tanks) or the Imaging Atmospheric Cherenkov Telescopes (IACTs). All $\gamma$-ray detectors are wide-field survey telescopes, except the IACTs.

\section{Sources of GWs and $\gamma$-rays}
NS mergers are important sources for both GW and $\gamma$-ray observatories, although they are not the only potential multimessenger sources.
Section \ref{sec:NSM} describes these mergers and the science they enable, while Section \ref{sec:other} discusses other possible multimessenger sources. Both sections discuss figures of merit for $\gamma$-ray observatories used in Table \ref{tab:gamma-rays}.

\subsection{Neutron Star Mergers}\label{sec:NSM}
BNS and (some) NSBH mergers, collectively referred to as NS mergers, produce short gamma-ray bursts (SGRBs) as well as kilonovae. 
These events have GW signals in the $\sim$Hz-kHz range and are found with CBC searches.
SGRBs are observed from $\sim$keV-MeV energies in their prompt phase. Prompt emission is followed by afterglow that has been observed from radio to GeV energies, and may soon be detected in the TeV regime based on the MAGIC detection of the long GRB 190114C \citep{MAGIC_190114C} and the sensitivity improvement with the CTA. As SGRBs arise from collimated jets \citep[e.g.][]{SGRBs_fong}, they are not expected to be detectable for the majority of NS mergers; GW radiation is omnidirectional but not isotropic \citep{GW_inclination}. Accounting for these effects, about 10-15\% of GW-detected NS mergers will produce SGRBs with Earth in the jet opening angle.

Joint GW-GRB detections of NS mergers give unique insights into relativistic jets, astroparticle physics, and the equation of state (EOS) of supranuclear matter, and provide precise tests of fundamental physics. A more in-depth summary of the science enabled by the multimessenger observations of these events is available in \citep{WP_NS_mergers}. Because SGRB prompt emission occurs within a few seconds of merger, this science requires serendipitous observations of mergers. Therefore, our first SGRB figure of merit is the average sky coverage, corresponding to the probability a merger will be observed. Our second figure of merit is the known or predicted rate of SGRB detections, which directly corresponds to the likelihood of joint detections. Lastly, joint searches for GWs and GRBs will result in more confirmed GWs, GRBs, and joint detections, and enable a near real-time combination of localization information. These capabilities aid the coordinated follow-up effort, helping with their use in cosmology as standard sirens (with GRBs breaking inclination-distance correlations), the origin of heavy elements, and a fuller understanding of the NS EOS \citep[see, e.g.][and references therein]{WP_NS_mergers}. Therefore, localization accuracy is our last SGRB figure of merit. Localization with $\gamma$-ray instruments can occur in two ways: autonomous real-time prompt SGRB localization by a single detector (which can be improved with follow-up by other instruments on the same spacecraft) or the detector's use in the InterPlanetary Network (IPN) \citep{IPN} for timing annulus localizations. We note that instruments in Low Earth Orbit (LEO) require distant instruments for these annuli to be constraining, given the limited timing accuracy from SGRB observations.

\subsection{Other Possible Joint Sources}\label{sec:other}
The eighth thematic area for Astro2020 is Multimessenger Astronomy and Astrophysics. This includes the identification of sources of GWs and $\gamma$-rays. Here, we list most of the putative GW sources and their expected $\gamma$-ray emission. Beyond identifying the sources themselves, such detections could give insight into the formation processes of NSs and BHs, the formation channels of the binaries, the NS EOS, and the evolutionary pathways of supermassive BHs and galaxy formation. The serendipitous joint observations of these two messengers can be the catalyst for coordinated follow-up, another part of the last Astro2020 thematic area. Multimessenger science may be key to identifying GW sources, by providing a known position, time of interest, or directly measuring frequency evolution. Joint sources that require long-term EM monitoring cannot be studied by scintillators, can be studied by coded masks and IACTs, and are best studied with Cherenkov, Compton, and Pair-conversion survey telescopes. For such sources our two figures of merit are (total) sky coverage and cadence.

\begin{itemize}[noitemsep,topsep=0pt]
    \item \textbf{Core Collapse Supernovae} in the Milky Way may produce detectable GW burst emission, and are expected to occur once every few decades. $\gamma$-rays in the $\sim$MeV range measure the production of radioactive elements that probe both stellar convection and the supernova engine. Some extreme CCSN also power long GRBs, which may produce significantly stronger GW emission \cite{LGRB_1,LGRB_2}. Neutrinos from collapse events are also detectable, and joint GW, neutrino, and $\gamma$-ray detections would constrain both our understanding of the supernova engine and the physics behind it \citep{WP_CCSN}, but to do so requires improved sensitivity to both the ground-based GW network and MeV $\gamma$-ray observatories.
    \item \textbf{Pulsars} are rapidly rotating neutron stars with large magnetic fields from which we observe pulsed electromagnetic emission. Any non-axisymmetric deformation in the object would cause it to emit $\sim$Hz-kHz CWs, generally at twice the rotational frequency \citep[see, e.g.][for a review]{GWs_isolated_NS}. $\gamma$-ray monitoring of pulsars provide accurate timing solutions that enable deep searches for CWs. Further, $\gamma$-ray observatories have also provided dozens of well-behaved pulsars for use in PTAs.
    \item \textbf{Accreting NSs} are promising sources of intermediate duration GWs. Small deformations may survive on short timescales at the regions of accretion \citep{Watts_1, watts_2}. $\gamma$-ray observations can measure the frequency change and inform on the accretion rate. A possible example are transitional pulsars which spin up during accretion but otherwise have normal spin-down behavior. 
    \item \textbf{Pulsar Glitches} are sudden changes in the rotation period and period derivative of the pulsar, with recovery timescales of hours to months. Glitches are thought to be caused by interactions at the core-crust interface, which could produce GWs during the recovery period. All-sky $\gamma$-ray monitors with localization capability can constrain the glitch time of $\gamma$-ray pulsars to an accuracy of $\sim$minutes, as was recently done for a Vela pulsar glitch \citep{Vela_Kerr}. This precision enables sensitive follow-up searches for intermediate-duration GW searches by providing both a start time and a known frequency evolution.
    \item \textbf{Giant Magnetar Flares} are short, bright flashes of $\gamma$ rays, followed by quasi-periodic oscillations (QPOs) for hundreds of seconds \citep[see, e.g.][]{GMF_Palmer,GMF_Hurley,watts_QPO}. They could result in non-axisymmetric deformations of the magnetar through crust-cracking or magnetic field-induced structural changes \citep{MF_1}. The prompt flare may produce GW burst emission, and the QPOs provide a known frequency to search for intermediate-duration GWs \citep{GMF_GWs,LVC_SGR}. Joint observations can inform on the NS structure and the emission mechanism for magnetar flares.
    \item \textbf{Supermassive Black Holes Binaries} (SMBHBs) and their mergers are key sources for PTAs and LISA. 
    The evolutionary pathways to creating SMBHBs are intricately tied to galaxy formation, but poorly understood. Long timescale observations of active galactic nuclei can reveal periodicity that may be related to future GW sources. The BL Lac object PG 1553+113 has an apparent 2.2 year cycle that has been observed for $\sim$5 periods by the Fermi LAT, which could arise from a SMBHB system \citep{ackermann2015multiwavelength,Tavani2018blazar}. Observations of MeV blazars would allow for multimessenger constraints on the formation of SMBHBs, without necessarily observing the same individual sources.
    \item \textbf{Something Unexpected.} Among the most interesting options for GW and multimessenger sources are those that we do not predict. One such example may be the creation of a SGRB following stellar-mass binary black hole mergers \citep[e.g.][]{Connaughton2016}, which is generally unexpected due to the lack of available matter. Because of the low rate of $\gamma$-ray transients and $\gamma$-ray detectors being all-sky monitors, they are promising EM partners for the unexpected, and could reliably prove association even without prior statistical assumptions.
\end{itemize}

\section{Summary}
Joint $\gamma$-ray and GW searches will identify the sources of both of these messengers and initiate the coordinated follow-up efforts. The all-sky monitoring capability of $\gamma$-ray and GW facilities, and the expected source types, make them synergistic partners in the multimessenger era.

The science possible with NS mergers is incredible. To ensure success, \textit{we specifically recommend a vigorous upgrade timeline for the ground-based GW network and continued and improved $\sim$keV-MeV $\gamma$-ray coverage. Further, we recommend the creation of a coherent GRB network analogous to the ground-based GW network.}  
From Table \ref{tab:GW_network}, small-scale missions can contribute critical sky coverage, sensitivity, and localizations for SGRB studies, though not at the same time. 
High energy missions that do not have GW counterparts as prime science drivers will provide important coverage and capabilities for joint $\gamma$-ray and GW science. 

To study longer-duration transients generally requires larger-scale missions that can study individual sources in detail. Any new joint detection will uncover another class of GW sources, and will enable unique science. To capture the full range of multimessenger sources, \textit{we recommend broad coverage of the $\gamma$-ray sky from keV to TeV energies in partnership with broad coverage of the GW spectrum}.

\pagebreak

\begin{landscape}
\begin{table}
\small
\begin{tabular}{|c|c|c|c|c||c|c|c||c|c|c|}
\hline
Mission/               & Mission & Start & $\gamma$-ray Detector;    & Energy         & Average & SGRB                              & Location                                                                        & Sky                                  &   &                                                                             \\
Instrument             & Class   & Year  & Other Coverage & Range          & View           & Rate                               & Accuracy                                                                            & View                               & Cadence        & Ref.                                                                \\ \hline
Swift/BAT              & MIDEX   & 2004- & Coded Mask; XUVO          & 15-150 keV     & 15\%               & 8.1                                    & $\sim$2'$^\dagger$                                                                  & 88\% & Daily       & \citep{BAT,BAT_GRB}  \\     
Fermi/GBM              & Probe   & 2008- & Scintillators; $\gamma$   & 8-40,000 keV   & 50\%               & 39.5$^1$                      & $\sim$12$^\circ$ &                                        &                & \citep{GBM,GBM_GRB} \\      
Fermi/LAT              & Probe   & 2008- & Pair conversion; $\gamma$ & 0.04-300 GeV & 20\%               & 1.7                                    & $\sim$0.5$^\circ$                                                                   & 100\%                                  & 3 Hours        & \citep{LAT,LAT_GRB}\\       
IACTS$^2$ &         & 2004- & Cherenkov (Air)           & 0.1-100 TeV    & Pointed            & ?                                      & \textless{}0.1$^\circ$                                                              &                                        &                & \citep{HESS_GW} \\          
HAWC                   &         & 2015- & Cherenkov (Water)         & 0.1-100 TeV    & 15\%               & ?                                      & $<$1$^\circ$                                                                              & 67\%                                   & Daily          & \citep{HAWC_1,HAWC_2}\\     \hline
BurstCube              & CubeSat & 2021  & Scintillators             & 30-1000 keV    & $>$30\%            & $>$20                                  & $\sim$20$^\circ$                                                                    &                                        &                & \citep{BurstCube}    \\      
SVOM/ECLAIRS           &         & 2021  & Coded Mask; $\gamma$XO    & 4-120 keV      & 15\%               & 7                                      & $<$14'                                                                              & $\sim$50\% & Variable       & \citep{SVOM}   \\            
SVOM/GRM               &         & 2021  & Scintillators; $\gamma$XO & 50-5000 keV    & 15\%               & 15                                     &                                                                                     &                                        &                & \citep{SVOM}               \\
Glowbug                & MoO     & 2023  & Scintillators             & 30-2000 keV    & 50\%               & 50                                     & $\sim$9-12$^\circ$                                                                  &                                        &                & \\                                             \hline
CTA                    &         & 2022  & Cherenkov (Air)           & 0.05-300 TeV   & Pointed            & ?                                      & $<$0.05$^\circ$                                                                     &                                        &                & \citep{CTA_1,CTA_2} \\     
SGSO                   &         & 2022  & Cherenkov (Water)         & 0.1-100 TeV    & 24\%               & ?                                      & $<$0.4$^\circ$                                                                      & 67\%                                   & Daily          & \citep{SGSO} \\              
Bia (2)                  & MoO     & 2025  & Scintillators             & 20-2000 keV    & 80\%               & 80-120                                 & $\sim$10-12$^\circ$                                                                 &                                        &                & \\                                            
COSI-SMEX              & SMEX    & 2025  & Compton                   & 100-5000 keV   & 25\%               & 10-20                                  & $\sim$0.5$^\circ$                                                                   & 80\%$^3$                             & 1.5 Hours & \\                                            
                      &         &       & Scintillators             &  150-5000 keV     & 50\%               & 10-20                                  &                                                                                     &                                        &                & \\                                            
MoonBEAM               & MoO     & 2025  & Scintillators             & 30-1000 keV    & 100\%              & 35-40                                  & $\sim$1$^\circ$$^4$                                                                  &                                        &                & \\                                            
Nimble/HAM             & SMEX    & 2025  & Scint.; UVOIR      & 20-3000 keV    & 40\%               & 25-45                                  & $\sim$12-15$^\circ$$^\dagger$                                                               &                                        &                & \\                                            
AMEGO                  & Probe   & 2030  & Compton, Pair             & 0.2-30,000 MeV & 20\%               & 60-100                                 & $\sim$0.5$^\circ$                                                                   & 100\%                                  & 3 Hours        & \citep{AMEGO_1,AMEGO_2} \\ 
STROBE-X               & Probe   & 2030  & Coded Mask; X             & 2-50 keV       & 30\%               & 7-10                                   & $\sim$2'                                                                            & 67\%                                   & Variable       & \citep{STROBEX}  \\
THESEUS                &         & 2032  & Coded Mask; XIR           & 2-20,000 keV   & 30\%               & 15-35                                  & 5'$^\dagger$                                                                        & 64\% & 4.5 Hours      & \citep{THESEUS}    \\         \hline
\end{tabular}

\caption{A summary of representative active (first block), funded (second block) and proposed $\gamma$-ray missions. Some $\gamma$-ray missions have instruments observing other wavelengths, which are abbreviated here as $\gamma$-ray ($\gamma$), X-ray (X), Ultraviolet (UV), Optical (O), and Infrared (IR). The double horizontal lines separate out the figures of merit for SGRB detection and surveys. The Average View column refers to the field of view modulo livetime, corresponding to the likelihood a given observatory will observe an event. The location accuracy is given in terms of 1-sigma uncertainties (statistical and systematic). Sky View and Cadence refer to how often that fraction of the sky is observed. We do not list estimated SGRB rates for the Cherenkov telescopes as they are unknown. For future missions these estimates are based on the current design of the mission parameters and instruments and are subject to change. Representative values are assumed when not precisely known.
$^\dagger$ denotes missions with on-board follow-up instruments that enable more accurate localizations.
$^1$GBM also has subthreshold searches, identifying an additional $\sim$80 SGRB candidates/year.
$^2$H.E.S.S./MAGIC/VERITAS.
$^3$COSI-SMEX observes 100\% of the sky with a daily cadence.
$^4$This localization assumes an IPN annulus in partnership with a LEO mission.}\label{tab:gamma-rays}
\end{table}
\end{landscape}


\pagebreak

\bibliographystyle{abbrv}
\bibliography{references}

\end{document}